\documentclass[aps,twocolumn,showpacs]{revtex4}
\usepackage{epsfig}

\begin{document}

\def\beq{\begin{equation}}
\def\eeq{\end{equation}}

\title{On the Physical Interpretation of Malyshkin's (2008) Model 
of Resistive Hall-MHD Reconnection}
\author{Dmitri A. Uzdensky}
\email{uzdensky@astro.princeton.edu}
\affiliation{Princeton University/ 
Center for Magnetic Self-Organization (CMSO), Princeton, NJ 08544}

\date{\today}

\begin{abstract}
A simple Sweet--Parker-like model for the electron current layer in 
resistive Hall magnetohydrodynamic (MHD) reconnection is presented,
with the focus on the collisionless limit. The derivation readily 
recovers the main results obtained recently by Malyshkin [PRL, {\bf 101}, 
225001 (2008)] and others, but is much quicker and more physically 
transparent.
In particular, it highlights the role of resistive drag in determining
the electron outflow velocity. The principal limitations of any such 
approach are discussed.
\end{abstract}

\pacs{52.30.Cv, 52.30.Ex, 52.35.Vd, 94.30.cp}

\maketitle


\section{Introduction}
\label{sec-intro}


Recently, Malyshkin \cite{Malyshkin-2008} proposed an analytical model 
of reconnection in resistive Hall MHD regime without electron inertia
(see also a recent paper by Simakov \& Cha\'con 2008, Ref.~\cite
{Simakov_Chacon-2008}).
Malyshkin's derivation was based on the local Taylor expansion of 
the resistive Hall MHD equations at the center of the electron layer. 
Under several simplifying assumptions, he was able to obtain expressions 
for the main parameters of the reconnection layer, including the reconnection 
rate~$E_z$, in terms of a few input parameters, namely, the length of the 
resistive electron layer~$L$, the magnetic field just outside the layer~$B_m$, 
the density~$n$, and the magnetic diffusivity~$\eta=\eta' c^2/4\pi$
(where $\eta'$ is the resistivity) \cite{Malyshkin-2008}. 
One of the main results was that the reconnection rate becomes 
independent of the resistivity in the limit~$\eta\rightarrow 0$
and scales inversely with~$L$.

In this contribution we show that, under the assumptions of Malyshkin's 
model, his main results can be obtained much quicker and easier, in a 
way that we believe is more transparent and lends itself more readily 
to a clear physical interpretation. 
Specifically, we point out that Malyshkin's (2008) expressions 
for the reconnection layer parameters follow straight-forwardly 
from (essentially) a Sweet--Parker-like analysis applied to the 
inner electron layer. To make a direct comparison with Malyshkin's 
paper~\cite{Malyshkin-2008} easier, we adopt the same system of 
coordinates that he used, i.e., $x$ is the direction across the 
layer, $y$ is the outflow direction along the layer, and $z$ is 
the ignorable direction.
The only major modification in the Sweet--Parker procedure that 
one needs to make for the problem under consideration, is in 
the (electron) equation of motion (generalized Ohm's law) in 
the outflow ($y$) direction.
Namely, since the electron mass is completely neglected here, 
the electron outflow velocity $v_{ey}$ is no longer controlled by 
their inertia, as it would be in the classical Sweet--Parker approach.
Instead, it is determined by the balance between the outward acceleration 
due to the Lorentz force (and perhaps a comparable contribution from the 
pressure gradient force) and the collisional (resistive) drag exerted on 
the electrons by the much slower moving ions:
\beq
{{B_x j_{ez}}\over c} \simeq -\, ne\, \eta' j_{ey} = n^2 e^2 \eta' v_{ey}
\ \Rightarrow \ v_{ey} \sim {{B_x j_{ez}}\over{cn^2e^2\eta'}}\, ,
\label{eq-v_ey}
\eeq
where the characteristic values for $v_{ey}$ and~$B_x$ are 
taken at the outflow end of the electron layer, $x=0,y=L$.

Next, using the $z$ component of the generalized Ohm law 
(the electron equation of motion) both at the end of the electron 
layer ($x=0,y=L$) and at the origin~$O$ ($x=0=y$), together with 
the stationarity condition $E_z(x,y) = {\rm const}$, we can estimate 
$B_x \sim  cE_z/v_{ey} = (c/v_{ey})\, \eta' j_{ez}$. Substituting
this into equation~(\ref{eq-v_ey}), we immediately obtain
\begin{eqnarray}
v_{ey} &\sim&  |v_{ez}| = j_{ez}/ne \sim c B_m/4\pi n e \delta \nonumber \\
&\Rightarrow& \ v_{ey} \delta \sim  {{cB_m}\over{4\pi n e}} 
= d_e V_{Ae} = d_i V_A \, ,
\label{eq-v_ey-delta}
\end{eqnarray}
consistent with the expectation that the out-of-plane 
magnetic field $B_z\sim B_m$ (e.g., Refs.~\cite{Uzdensky_Kulsrud-2006,
Yamada_etal-2006}). Note that the resistivity has dropped out, because 
similar resistive terms enter in both the~$z$ and~$y$ components of the 
electron equation of motion, balancing the electric force $-\,ne\,E_z$ 
and the Lorentz force $B_x j_{ez}/c$, respectively. 

Combining equation~(\ref{eq-v_ey-delta}) with the two remaining 
relationships of the standard Sweet--Parker analysis: the incompressibility 
condition, $|v_{ex}|\,L = v_{ey}\,\delta$, and the $z$-component of Ohm's law
considered at the origin~$O$ and at the point~$M=(x=\delta,y=0)$ just above 
the resistive layer, 
$\eta B_m/\delta = cE_z = |v_{ex}| B_m \Rightarrow |v_{ex}|\,\delta  =\eta$,
we immediately get 
\begin{eqnarray}
\delta &\sim & {{\eta L}\over{d_i V_A}} = {L^2\over{Sd_i}} \, , \\
|v_{ex}| &\sim & {d_i\over L}\, V_A \, , \\
v_{ey} &\sim & |v_{ez}| \sim {{d_i V_A}\over{\delta}} \sim
V_A\, {{S d_i^2}\over{L^2}}\, ,  \\
c\,E_z &\sim & {d_i\over L}\, V_A\, B_m  \, ,
\label{eq-rec-rate}
\end{eqnarray}
which coincide with the corresponding Malyshkin's \cite{Malyshkin-2008}
expressions (20)-(26). Here, $S\equiv LV_A/\eta$ is the Lundquist number.
Note that $E_z$ is actually independent of~the ion mass~$m_i$. In addition, 
it is also independent of~$\eta$, and the fundamental reason for this is 
that resistive drag controls the electron flow velocity in both the~$z$ 
and~$y$ directions in a similar way.

We note that recently similar results in resistive Hall-MHD were 
independently obtained by Simakov \& Chac\'on~\cite{Simakov_Chacon-2008}, 
and even earlier by Wang~et~al.~\cite{Wang_etal-2000}. Furthermore, 
the expression~(\ref{eq-rec-rate}) for the collisionless reconnection 
rate dates at least as far back as Ref.~\cite{Cowley-1985} 
(Malyshkin 2008, private communication).

Finally, we would like to stress that neither 
Malyshkin~\cite{Malyshkin-2008}, nor Simakov \& 
Chac\'on~\cite{Simakov_Chacon-2008} address what, 
to us, seems to be the most important question in 
collisionless reconnection research: what determines 
the length~$L$ (in the outfow direction~$y$) of the 
electron and ion layers? This question is similar to 
the old question about the length of the central diffusion 
region in the Petschek reconnection model.
Both Malyshkin~\cite{Malyshkin-2008} and Simakov \& 
Chac\'on~\cite{Simakov_Chacon-2008} consider~$L$ (or $w$ 
in Simakov \& Chac\'on's notation) just an input parameter 
whose determination is beyond the scope of their papers. 
In our view, however, the question of whether the layer is macroscopic 
(independent of the local plasma parameters such as~$\eta$, $d_i$, $d_e$, 
etc.) or microscopic, is of great importance, both fundamental and
practical, since, for most space- and astrophysical systems, the 
global system size is usually by many orders of magnitude larger 
than~$d_i$, say.
Thus, if $L$ is comparable to the global system size, the resulting 
reconnection rate~(\ref{eq-rec-rate}), even though independent of the 
resistivity, is too slow to explain many observable phenomena, such as 
solar flares. On the other hand, if $L$ is microscopic (as seems to be 
indicated by numerical simulations~\cite{Shay-1998, Cassak-2005}; 
see, however, Refs.~\cite{Wang_etal-2000,Daughton_etal-2006}), 
it should be determined self-consistently as a part of the local analysis.
Thus, until the issue of the length of the reconnection layer is resolved, 
achieving a final complete theory of collisionless reconnection cannot be 
claimed. 

Likewise, neither the present analysis, nor the papers 
by Malyshkin~\cite{Malyshkin-2008} and Simakov \& Cha\'con
\cite{Simakov_Chacon-2008}, attempt to estimate the magnetic
field~$B_m$ just outside the electron layer in terms of the
global reconnecting magnetic field~$B_0$ measured just outside 
the ion layer (R.~Kulsrud 2008, private communication). 
In general, depending on the strength of the electron and
ion currents in the ion diffusion region ($\delta_e\ll x \ll \delta_i$),
$B_m$ may be much smaller than $B_0$ or comparable to~$B_0$.
Until this issue is resolved, the solution of the reconnection
problem remains incomplete.


\begin{acknowledgments}

I am very grateful to R.~Kulsrud, M.~Yamada, and L.~Malyshkin 
for valuable discussions.

This work is supported by National Science Foundation Grant 
No.~PHY-0215581 (PFC: Center for Magnetic Self-Organization 
in Laboratory and Astrophysical Plasmas).

\end{acknowledgments}



\end{document}